\documentclass[12pt]{article}

\pdfoutput=1
\usepackage{graphicx}
\usepackage{amsmath}
\usepackage{amsfonts}
\usepackage{amssymb}

\makeatletter
    \def\thebibliography#1{\noindent \textsc{\large References}\@mkboth
      {REFERENCES}{REFERENCES}\list
      {[\arabic{enumi}]}{\settowidth\labelwidth{[#1]}\leftmargin\labelwidth
	\advance\leftmargin\labelsep
	\usecounter{enumi}}
	\def\newblock{\hskip .11em plus .33em minus .07em}
	\sloppy\clubpenalty4000\widowpenalty4000
	\sfcode`\.=1000\relax}
    \makeatother

\begin{document}

\begin{flushleft}
\textbf{$Z'$, $Z_{KK}$, $Z^*$ and all that: current bounds and theoretical 
prejudices on heavy neutral vector bosons.}

\vspace{30pt}
\textsc{R.Contino} 

\vspace{8pt}
\textit{\small CERN Physics Department, Theory Division, CH-1211 Geneva 23, Switzerland}

\vspace{30pt}
\textbf{Abstract.} --
I review the current experimental bounds and theoretical predictions
for different kinds of heavy neutral vector bosons.

\end{flushleft}
\vspace{40pt}

\noindent \textbf{1. -- The case of  a ``standard'' $Z'$}
\vspace{10pt}

Heavy neutral vector bosons appear in the particle content
of many extensions of the Standard Model (SM), and their detection
in dilepton channels is often mentioned
as one example of early discovery mode at the Large Hadron Collider (LHC).
I this talk I will briefly review the theoretical lore and the present experimental
bounds on these heavy particles.

The interaction of a heavy 
$Z'$ with the Standard
Model fermions and Higgs boson can be parametrized in terms of the overall coupling strengths 
$g_{Z'}^{[\Psi]}$, $g_{Z'}^{[H]}$, and of (quantized) charges $z_a$, $z_H$ :
\begin{equation}
{\cal L}_{Z'} = -\frac{1}{4} Z_{\mu\nu}^\prime Z^{\prime\,\mu\nu} + g_{Z'}^{[\Psi]} \sum_a Z'_\mu \, 
 \bar\Psi_a z_a \gamma^\mu \Psi_a + g_{Z'}^{[H]}\, H^\dagger z_H Z'_\mu iD^\mu H + h.c. \, ,
\end{equation}
where $a=Q,d,u,L,e$ runs over all SU(2)$_L$ SM representations. The last term, in particular, leads to 
$Z'-Z$ mixing after electroweak symmetry breaking (EWSB).

A natural possibility is that the $Z'$ is the gauge field of a
new local symmetry broken at high energies, so that $g_{Z'}^{[\Psi]}=g_{Z'}^{[H]}=g_{Z'}$.
To avoid large flavor-changing neutral currents, it is also commonly assumed in the literature 
that the couplings of the $Z'$ to the SM fermions are flavor universal.
I will refer to this case as that of a ``standard''~$Z'$~\cite{Langacker:2008yv}. 
Examples are: a heavy $B-L$ gauge boson; a $Z_\psi$ from $E_6$ Grand Unified Theories, 
where $E_6\to$ SO(10)$\times$ U(1)$_\psi$ at the unification scale ($g_{Z'}=g_2 \tan\theta_W$ and charge equal 
to $\sqrt{5/72}$ ($-\sqrt{5/72}$) for all left-handed (right-handed) SM fermions); 
a heavy replica of the SM $Z$,
($g_{Z'}=g_2/\cos\theta_W$ and 
charges given by $(T_{3_L}-Q\sin^2\theta_W)$);  a heavy replica of the hypercharge.

\begin{figure}[t!]
\begin{minipage}{0.32\linewidth}
\includegraphics[width=4.8cm]{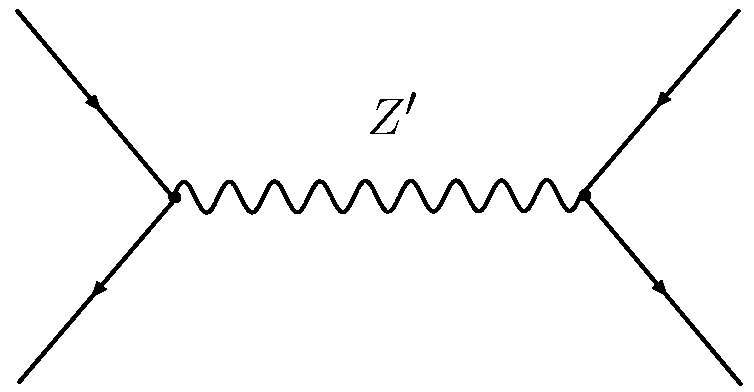}
\end{minipage} 
$\propto \displaystyle \frac{g_{Z'}^{[\Psi]\, 2}}{M_{Z'}^2}$ \\[-0.2cm]
\hspace*{2.2cm} (a)
\hspace{0.5cm} \\[0.8cm]
\begin{minipage}{0.285\linewidth} \vspace*{-0.36cm} 
\includegraphics[width=4.1cm]{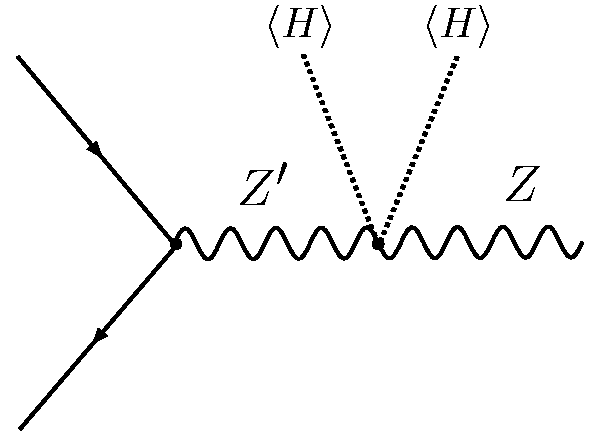}
\end{minipage} 
$\propto \displaystyle \frac{g_{Z'}^{[H]}g_{Z'}^{[\Psi]}}{M_{Z'}^2}$
\hspace{0.8cm}
\begin{minipage}{0.33\linewidth} \vspace{-1.35cm}
\includegraphics[width=4.7cm]{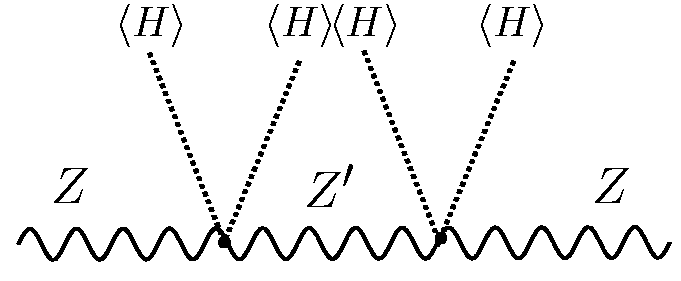} 
\end{minipage} 
$\propto \displaystyle \frac{g_{Z'}^{[H]\, 2}}{M_{Z'}^2}$ \\[-0.2cm]
\hspace*{2.2cm} (b) \hspace{6.5cm} (c)
\caption{\label{fig:EWdiagrams} \small
Corrections to LEP observables from the tree-level exchange of a heavy $Z'$:
contact interactions (a); vertex corrections (b); corrections to the $\rho$
parameter (c).
}
\end{figure}

The main indirect bounds on the $Z'$ mass and couplings come from LEP experiments.
A tree-level exchange of a heavy $Z'$ affects both the off-pole LEP2 observables,
through the generation of four-fermion contact interactions $\propto g_{Z'}^{[\Psi]\, 2}$,
and the $Z$-pole LEP1 observables via the $Z'-Z$ mixing, see figure~\ref{fig:EWdiagrams}. 
In the case of the latter, it is useful to distinguish between vertex corrections 
$\propto g_{Z'}^{[H]} g_{Z'}^{[\Psi]}$, 
and corrections to the $\rho$ parameter $\propto g_{Z'}^{[H]\, 2}$.
These corrections are said to be oblique, or universal, if they can be recast
as pure modifications of the SM gauge boson self energies by making a suitable redefinition of fields.
In this case it has been shown that all LEP observables can be expressed in terms of
only four parameters, or form factors: $\hat S$, $\hat T$, $W$, $Y$~\cite{Barbieri:2004qk}.
For example, a heavy hypercharge leads to oblique corrections, while a heavy $B-L$,
a $Z_\psi$ or a heavy $Z$ are not oblique. Notice that the oblique basis does not usually coincide with
the mass-eigenstates basis.
Besides LEP indirect constraints, direct exclusion limits on the Drell-Yan production 
$q\bar q \to Z' \to l^+ l^-$, $l=e,\mu$ come from D$\not$O and CDF experiments~\cite{tev}.

\begin{figure}[t!]
\begin{center}
\includegraphics[width=7.0cm]{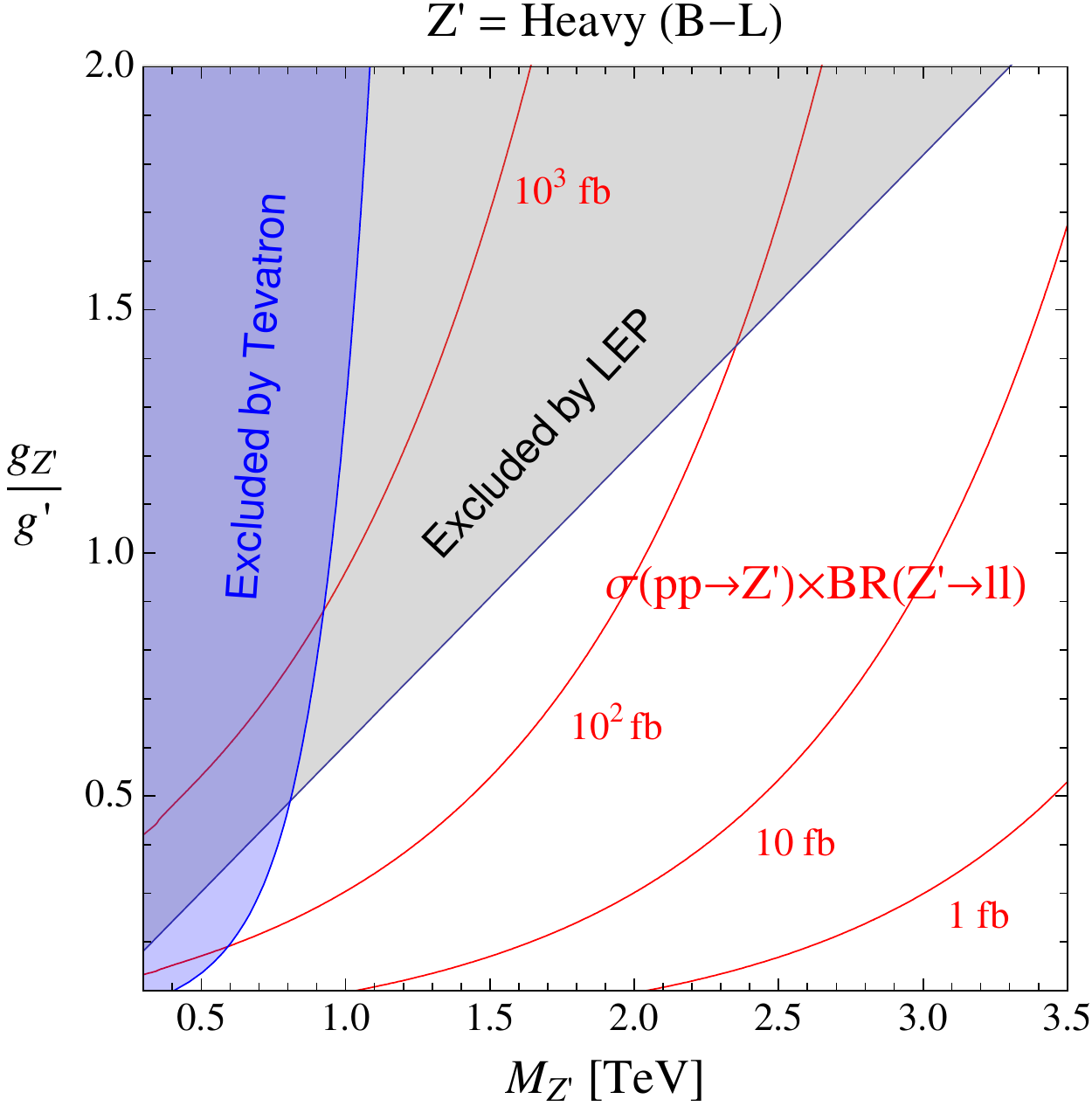}
\hspace{0.65cm}
\includegraphics[width=7.0cm]{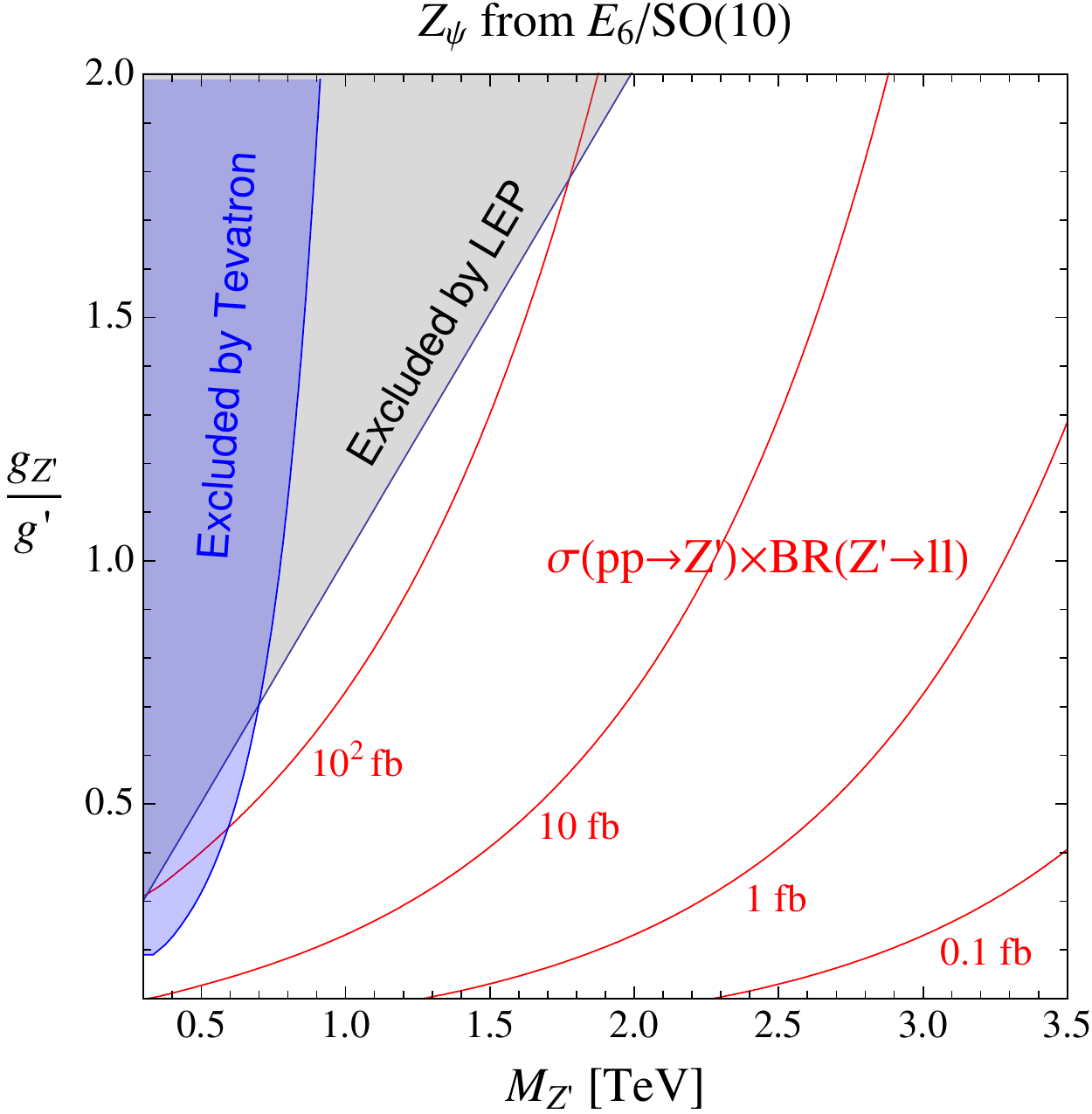}
\end{center}
\caption{\label{fig:heavyZ} \small  
Constraints on the mass and the coupling strength of a heavy $B-L$ (left plot) 
and a $Z_\psi$ from $E_6/$SO(10) (right plot). The coupling strengths are expressed in units
of the SM value $g'=g_2\tan\theta_W$. 
The red curves report the production rate $\sigma (pp\to Z' \to l^+l^-)$, with $l=e,\mu$.
}
\end{figure}
\begin{figure}[t!]
\begin{center}
\includegraphics[width=7.0cm]{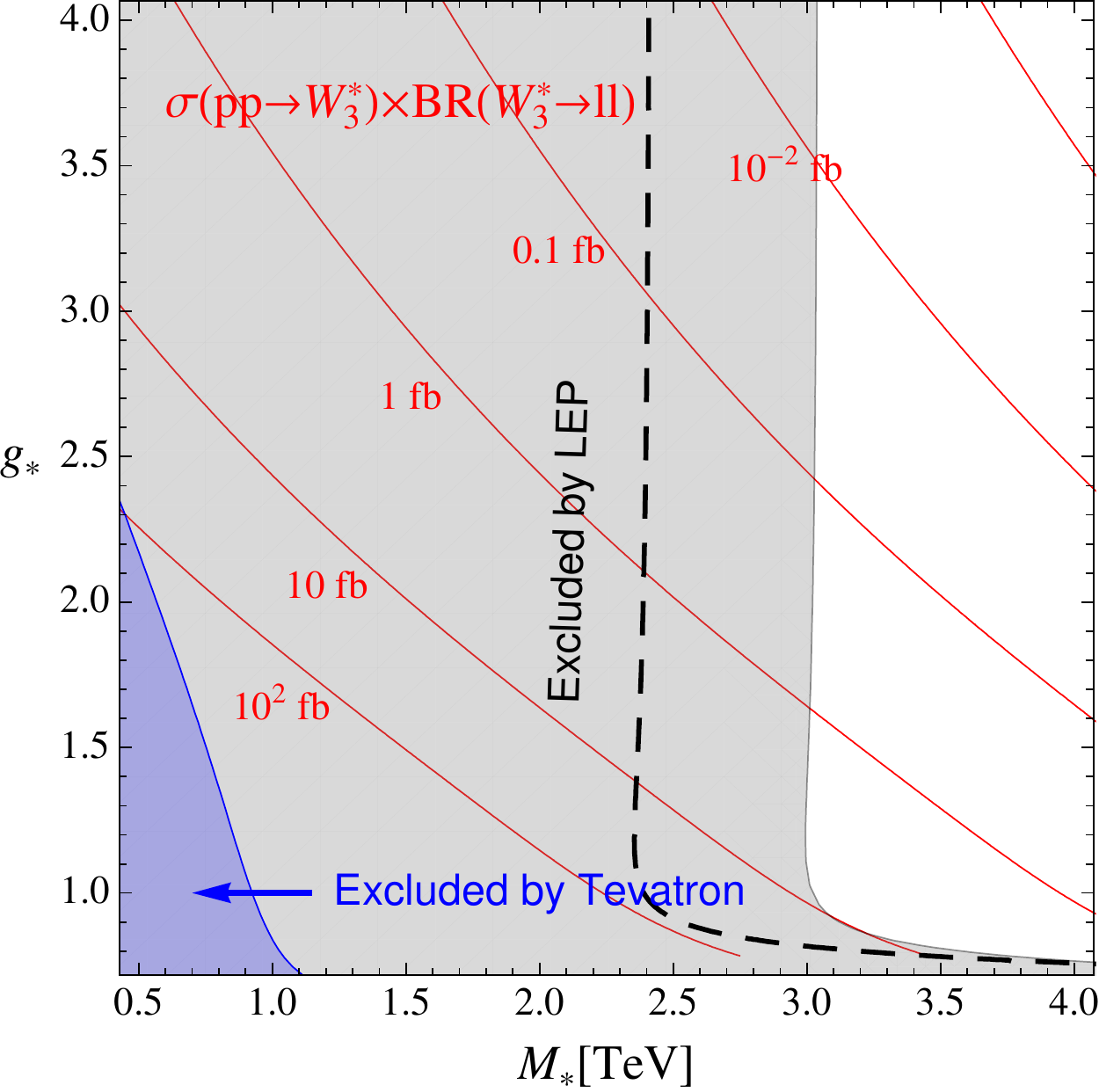}
\vspace*{-0.3cm}
\end{center}
\caption{\label{fig:TS}  \small
Constraints on the mass and the coupling strength of
vectorial resonances in the adjoint of SU(2)$_L\times$SU(2)$_R\times$U(1)$_{B-L}$ (see text).
The red curves report the production rate of the neutral $W^*_3$ to two charged leptons:
$\sigma (pp\to W^*_3 \to l^+l^-)$, with $l=e,\mu$.
}
\end{figure}

The left plot of figure~\ref{fig:heavyZ} shows the bounds on a heavy $B-L$  with mass
$M_{Z'}$ and arbitrary coupling strength $g_{Z'}$. 
The (darker) blue region is excluded by the CDF RunII results of~\cite{tev}, which were derived collecting 
1.3 fb$^{-1}$ of data.
The exclusion region in (lighter) gray is
instead obtained using the fit to the LEP electroweak data of 
reference~\cite{Cacciapaglia:2006pk}.~\footnote{The plots
in figure~\ref{fig:heavyZ} and~\ref{fig:TS} have been obtained computing the $Z'$ production 
cross section with the approximate formula in eq.(3.16) of reference \cite{Leike:1998wr}.
This gives a reasonably accurate result
except for low masses, where it slightly underestimates the cross section.}

For a $Z'$ coupling strength equal to the SM value, $g_{Z'}=g' = g_{2} \tan\theta_W$, 
CDF excludes masses up to $\sim 950\, \text{GeV}$, while the bound from LEP is $M_{Z'}\gtrsim 1.65\, \text{TeV}$.
Even for such latter values of the mass, the Drell-Yan production rate $pp\to Z' \to l^+l^-$ ($l=e,\mu$)
at the LHC is still large, $\sigma\times BR \simeq 200\, \text{fb}$, corresponding to a discovery luminosity 
of $\sim 100\, \text{pb}^{-1}$, see figure~\ref{fig:heavyZ} and~\cite{delreventura}.
Similar results hold for a $Z_\psi$, a heavy hypercharge and a heavy $Z$.
The case of the $Z_\psi$ is shown in the right plot of figure~\ref{fig:heavyZ}.

Larger values of $g_{Z'}$ are more constrained by LEP, requiring
larger integrated luminosities for a discovery. In particular, a strongly coupled $Z'$ ($1 \ll g_{Z'} < 4\pi$) 
is excluded by LEP unless it has a very large mass.
The most robust constraint comes from the contact interactions of fig.\ref{fig:EWdiagrams}(a), 
which are proportional to $g_{Z'}^{[\Psi]\, 2}$: for generic charges $z_a \sim {\cal O}(1)$ the ratio 
$(g_{Z'}^{[\Psi]}/M_{Z'})$ cannot be too large.
On the other hand, the tree-level correction to the $\rho$ parameter will vanish  if the $Z'$ originates 
from a new sector that preserves a custodial symmetry after EWSB~\cite{Sikivie:1980hm}.
In this case, the only other constraint comes from the vertex correction of fig.\ref{fig:EWdiagrams}(c), 
leaving the possibility of a large value of $g_{Z'}^{[H]}$ provided $g_{Z'}^{[\Psi]}$ is small.
In fact, different coupling strengths $g_{Z'}^{[H]}$ and $g_{Z'}^{[\Psi]}$ can naturally emerge in theories 
where both the heavy vector and the Higgs field are bound states of a new strongly-interacting dynamics.

\vspace{20pt}

\noindent \textbf{2. -- Strongly coupled  $Z^*$}
\vspace{10pt}

Particularly attractive and theoretically motivated is the possibility that the SM gauge fields themselves 
have some degree of compositeness: they could arise as admixtures of an elementary field $A_\mu$
with a tower of resonances of the strong sector, $\hat\rho_\mu$. In complete analogy with the
photon-$\rho$ mixing occurring in QCD, 
the $A_\mu-\hat\rho_\mu$ mixing can proceed via a mass term
\begin{equation}
{\cal L}_{mix} = \frac{M_*^2}{2} \left( \frac{g_{el}}{g_*} A_\mu - \hat\rho_\mu \right)^2 \, , 
\end{equation}
where $g_{el}$ and $g_*$ denote respectively the gauge coupling of $A_\mu$ and the coupling strength among three
composite states. It is natural to assume $1\ll g_* < 4\pi$, and $g_{el}\lesssim 1$.
Due to ${\cal L}_{mix}$, the combination $(g_{el} A_\mu - g_* \hat\rho_\mu)$ acquires a mass $M_*$,
while the orthogonal one -- to be identified with the SM gauge field -- remains massless.
After rotating from the elementary/composite basis to the mass-eigenstates basis, one has:
\begin{equation} \label{mixing}
\begin{split}
|\text{SM} \rangle =& \cos\theta\, |A_\mu\rangle + \sin\theta\, |\hat\rho_\mu \rangle \\[0.15cm]
|\text{heavy} \rangle =& -\sin\theta\, |A_\mu\rangle + \cos\theta\, |\hat\rho_\mu \rangle 
\end{split} \qquad\quad
\tan\theta = \frac{g_{el}}{g_*} \, , \quad
g_\text{SM} = \frac{g_{el}g_*}{\sqrt{g_{el}^2+g_*^2}} \simeq g_{el}\, .
\end{equation}
Here $\theta$ parametrizes the degree of compositeness of the SM field.
While the Higgs in such a scheme is a full composite (solving in this way the SM hierarchy problem via dimensional 
transmutation), the SM fermions can also be partial composites as a consequence of their mixing with a tower of 
fermionic resonances of the strong sector. 
Theories with a warped extra dimension, with their 4-dimensional dual description, are explicit realizations
of this scenario (see, for example,~\cite{Contino:2006nn}).
In particular, the tower of heavy resonances in the mass eigenstate basis is interpreted as the Kaluza-Klein (KK) modes
of the higher-dimensional theory.

The phenomenology of a heavy neutral vector $Z^*$ that is the result of the elementary/composite mixing of
eq.(\ref{mixing}) -- or, equivalently, that of a KK in a warped extra-dimensional
theory -- is rather different from the phenomenology of a ``standard'' $Z'$.
The LEP2  precision tests imply that the degree of compositeness of the light fermions
has to be very small. On the other hand, the large top quark mass can find a natural explanation if the top is
mainly composite.
This means that the Drell-Yan production of $Z^*$ will mostly proceed via the interaction of its elementary 
component with the light  partons of the proton. The coupling strength for this scattering is 
$ \sim g_{el}\sin\theta = g_\text{SM} \tan\theta$, implying a suppression factor $\tan\theta = g_{el}/g_* \ll 1$ compared to 
the SM value. Moreover, the same factor will also suppress the decay rate to light quarks and leptons.
The $Z^*$ will instead couple strongly, via its composite component, to all the SM particles with a sizable degree of 
compositeness: the Higgs, the longitudinally polarized $W$ and $Z$'s, the top and bottom quarks.
In other words, the $Z^*$ will mainly decay to $h Z_L$, $W^+_L W^-_L$, $t\bar t$ or $b\bar b$, while 
the branching ratio to a pair of electrons or muons will be quite small.
For example, the decay rates of a heavy $W^*_{3}$ (the neutral component of a heavy SU(2)$_L$ triplet) 
are~\cite{Contino:2006nn}
\begin{equation}
\begin{split}
& \Gamma\left( W^*_3 \to q\bar q\right) = 3\, \Gamma\left( W^*_3 \to l\bar l\right) 
  \simeq \frac{g_2^2 \tan^2\!\theta}{32\pi} M_* \\
& \Gamma\left( W^*_3 \to t_L\bar t_L\right) =\Gamma\left( W^*_3 \to b_L\bar b_L\right)  =
  \left( \sin^2\!\varphi_{L} \cot\theta - \cos^2\!\varphi_L \tan\theta \right) \frac{g_2^2}{32\pi} M_* \\
& \Gamma\left( W^*_3 \to Z_L h \right) = \Gamma\left( W^*_3 \to W^+_L W^-_L \right)  
  = \frac{g_2^2 \cot^2\!\theta}{192\pi} M_*\, ,
\end{split}
\end{equation}
where $\sin\varphi_L$ is the degree of compositeness of $t_L$ and $b_L$. For $M_*= 3\, \text{TeV}$, 
$\tan\theta = 1/6$, $\sin\varphi_L = 0.4$, one has a total decay rate $\Gamma_{tot}(W_3^*) = 170\,\text{GeV}$ and
$BR(ee , \mu\mu) = 0.1\%$, $BR(tt, bb) = 5\%$.

The plot of figure~\ref{fig:TS} shows the experimental constraints on a strong sector with 
an SU(2)$_L\times$SU(2)$_R\times$U(1)$_{B-L}$ global invariance and vectorial resonances
(two charged and three neutral states) in the adjoint representation.

The larger symmetry is required to have an unbroken custodial invariance after EWSB
and  protect the $\rho$ parameter from large tree-level corrections. 
At the same time, however, it leads to several massive vectors, which makes 
the constraint from the vertex corrections in fig.\ref{fig:EWdiagrams}(c) stronger. 
The (lighter) gray region represents the portion of the 
$(M_*,g_*)$ plane excluded by the LEP precision measurements. 
The LEP2 constraint on contact interactions becomes negligible for $g_* \gtrsim 1$,
since the coupling of the heavy vectors to the light SM fermions scales as 
$g_{Z^*}^{[\Psi]} = g_{SM} \tan\theta \simeq g_{SM} (g_{SM}/g_*)$.
In that limit the dominant constraint comes from the $\hat S$ parameter, which does not
depend on $g_*$ since $g_{Z^*}^{[H]}=g_{SM} \cot\theta \simeq g_{SM} (g_*/g_{SM})$, hence 
$g_{Z^*}^{[\Psi]}g_{Z^*}^{[H]} = g_{SM}^2 =$ constant.
The dashed curve shows how much the LEP bound is relaxed by adding a correction 
$\Delta\rho = \Delta \hat T = 1\times 10^{-3}$,
which might come, for example, from the 1-loop contribution of new heavy fermions. The blue area 
is the region excluded by the CDF data of reference~\cite{tev}.

Superimposed in red are the curves reporting the production rate of the neutral $W^*_3$ in the
dilepton channel. One can see that in most of the region allowed by LEP the rate is quite small,
so that the observation of this channel will probably require very large integrated luminosities.
On the other hand, a discovery can come earlier from the other final states. 
According to the analysis of reference~\cite{Agashe:2007ki}, a $Z^*$ with mass $M_*=2\,\text{TeV}\, (3\,\text{TeV})$ 
and $g_*\simeq 3.9$ can be discovered in the $Zh$ channel with $\sim 100 \,\text{fb}^{-1}\, (1 \,\text{ab}^{-1})$.
Observing the $W^+W^-$ channel will instead require more integrated luminosity.

\begin{center} * \hspace{0.2cm} * \hspace{0.2cm} *  \end{center}
I would like to thank Slava Rychkov and Alessandro Strumia for interesting discussions and useful comments.

\end{document}